\documentclass[prl,aps,twocolumn,showpacs,superscriptaddress]{revtex4}
\usepackage{amsmath}
\usepackage{amsfonts}
\usepackage{amssymb}
\usepackage{graphicx}
\usepackage{bm}

\begin{document}

\title{The London penetration depth in single crystals of Ba(Fe$_{1-x}$Co$%
_{x}$)$_{2}$As$_{2}$ at various doping levels}
\author{R.~T.~Gordon}
\author{C.~Martin}
\author{H.~Kim}
\author{N.~Ni}
\author{M.~A.~Tanatar}
\author{J.~Schmalian}
\affiliation{Ames Laboratory and Department of Physics \& Astronomy, Iowa State
University, Ames, Iowa 50011}
\author{I.~I.~Mazin}
\affiliation{Code 6393, Naval Research Laboratory, Washington, D.C. 20375}
\author{S.~L.~Bud'ko}
\author{P.~C.~Canfield}
\author{R.~Prozorov}
\email[corresponding author: ]{prozorov@ameslab.gov}
\affiliation{Ames Laboratory and Department of Physics \& Astronomy, Iowa State
University, Ames, Iowa 50011}
\date{18 December 2008}

\begin{abstract}
The London penetration depth $\lambda(T)$ has been measured in single
crystals of Ba(Fe$_{1-x}$Co$_{x}$)$_{2}$As$_{2}$ using the tunnel diode
resonator technique. The measured doping levels of $x=$~0.038, 0.047, 0.058, 0.074 and 0.10 range from
underdoped to overdoped concentrations. The measurements have shown that the density of carriers participating in superconductivity decreases sharply in the underdoped regime, but the penetration depth as a function of temperature exhibits a robust power law, $\Delta\lambda(T)\sim T^{n}$, for all measured dopings, with $n$ between 2 and 2.5. We discuss the implications of these results and possible interpretations of such robust behavior.
\end{abstract}

\maketitle

%74.25.Nf 	Response to electromagnetic fields (nuclear magnetic resonance, surface impedance, etc.)
%74.20.Rp 	Pairing symmetries (other than s-wave)
%74.20.Mn 	Nonconventional mechanisms (spin fluctuations, polarons and bipolarons, resonating valence bond model, anyon mechanism, marginal Fermi liquid, Luttinger liquid, etc.)

The structure and symmetry of the superconducting order parameter is of crucial importance for determining the pairing mechanism in the newly discovered Fe-based pnictide superconductors. A useful method of probing the gap structure
is to measure the magnetic penetration depth in single crystals. The two parent systems for which a considerable amount of effort has been put forth to study are REFeAsO (1111) and AEFe$_{2}$As$_{2}$ (122), where RE is a rare earth
and AE is an alkali earth.

In the fluorine doped, or oxygen deficient, 1111 system, the majority of experiments indicate a fully-gapped Fermi surface (FS). Measurements of the London penetration depth, $\lambda(T)$, using a tunnel diode resonator (TDR)
technique on NdFeAsO$_{0.9}$F$_{0.1}$ \cite{Martin2008} and SmFeAsO$_{1-x}$F$_{y}$ \cite{Malone2008} as well as microwave cavity perturbation on PrFeAsO$_{1-y}$ \cite{Hashimoto2008} have found an exponential temperature
dependence of $\lambda(T)$ at low temperatures. Similar conclusions have been reached by muon spin relaxation ($\mu$SR) studies and most point contact Andreev Reflection (PCAR) measurements \cite{Chen2008}, although
some of them have been interpreted in terms of nodal gaps. \cite{Wang2008}. Knight shift measurements indicate spin
singlet superconductivity. The spin-lattice relaxation rate, $1/T_{1}\sim T^{3}$, is characteristic of nodal superconductivity, but can be reconciled with the extended $s_{\pm}$ model by pair-breaking scattering
under particular assumptions about its strength and defect concentration \cite{Parker2008}.

\begin{figure}[b]
\includegraphics[width=.94\columnwidth]{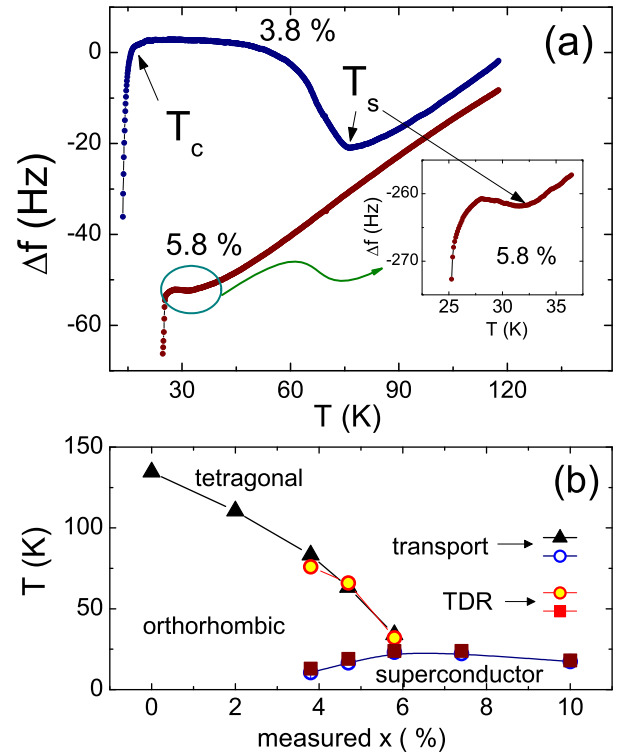}
\caption{(Color online) (a) Raw data for $x=$~0.038 (underdoped) and $x=$~0.058 (near optimal doping, the data has been
divided by a factor of 5 for clarity). The inset emphasizes a magnetic/structural transition. (b) phase diagram showing structural, $T_{s}$, and superconducting, $T_{c}$, transitions determined from transport \protect\cite{Nini2008} and TDR measurements.}
\label{fig1}
\end{figure}

The situation in the 122 system is somewhat more controversial. TDR measurements on Ba(Fe$_{0.93}$Co$_{0.07}$)$_{2}$As$_{2}$ have shown a clear non-exponential behavior of $\lambda(T)$ \cite{RTGordon2008}, whereas microwave measurements on Ba$_{1-x}$K$_{x}$Fe$_{2}$As$_{2}$ were interpreted
in terms of two fully opened superconducting gaps \cite{2Hashimoto2008}. In this paper, we focus on penetration depth studies of large single crystals of Ba(Fe$_{1-x}$Co$_{x}$)$_{2}$As$_{2}$ with different Co
dopings. We find that the penetration depth exhibits a robust power law, $\Delta\lambda(T)\approx C(T/T_{c})^{n}$, for all $x$. There is a clear change of regime at $x \sim 0.06$, where (i) the orthorhombic/antiferromagnetic -- tetragonal/nonmagnetic phase boundary crosses the superconducting phase boundary (Fig. \ref{fig1}), (ii) $n$ changes from 2.0$\pm0.1$ to 2.4$\pm$0.1 and (iii) the coefficient $C$ suddenly drops by an order of magnitude. This strongly suggests that the values of the exponent, and probably the power law itself, are not due to impurities, unless there is a sudden change in the impurity scattering around $x=$~0.06, but is an intrinsic characteristic, likely related to the proximity to the magnetic ordering/structural transition. Our separate study shows a $\lambda(T)\sim T^{2}$ behavior in a hole-doped 122 system, (Ba$_{1-x}$K$_{x}$)Fe$_{2}$As$_{2}$, as well.

Single crystals of Ba(Fe$_{1-x}$Co$_{x}$)$_{2}$As$_{2}$ were grown out
of self flux \cite{Nini2008}. The actual cobalt concentration
was determined by wavelength dispersive x-ray spectroscopy in the electron probe microanalyzer of a JEOL
JXA-8200 Superprobe. Magneto-optical imaging has revealed homogeneous Meissner
screening. Slabs with sizes of $\sim 1 \times 1 \times 0.2$ mm$^3$ and mirror-like
surfaces were cleaved with a razor blade from larger crystals.

\begin{figure}[tb]
\includegraphics[width=.95\columnwidth]{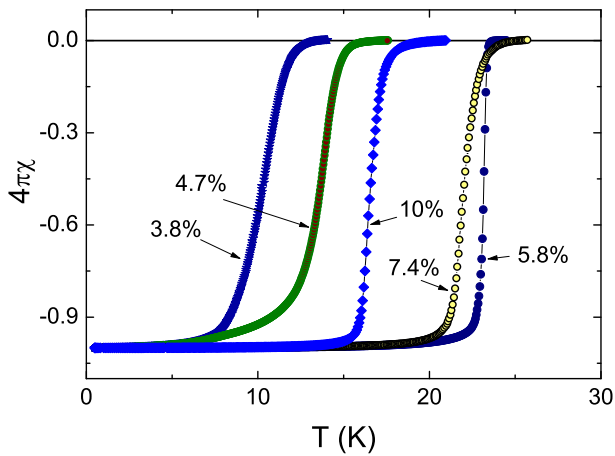}
\caption{(Color online) $4\pi\chi\left( T\right)$ in single crystals of
Ba(Fe$_{1-x}$Co$_{x}$)$_{2}$As$_{2}$ for different $x$.}
\label{fig2}
\end{figure}

The in-plane London penetration depth, $\lambda(T)$, has been measured using
a self-oscillating tunnel diode resonator (TDR) \cite{VanDegrift1975}. A sample to be studied is mounted on a sapphire rod and inserted into the inductor coil of the tank circuit. Throughout the measurement the temperature of the circuit is stabilized at 5.00 K $\pm$ 0.01 K. This is essential for stability in
the measured resonant frequency, which is resolved to about 0.01 Hz. This
translates to the ability to detect changes in $\lambda(T)$ in the range of
an \AA . The ac magnetic excitation field in the coil is about 20 mOe, which
is small enough to ensure that no vortices are present. The sample, with
magnetic susceptibility $\chi(T)$, leads to a change in the resonant
frequency of $\Delta f \equiv f(T)-f_{0} =-G 4 \pi\chi(T)$, where $f_{0}=1/2\pi\sqrt{LC} \approx 14$ MHz and $G\simeq
f_{0}V_{s}/2V_{c}\left( 1-N\right) $ is a geometrical calibration factor defined by the coil characteristics 
and the sample volume $V_{s}$. $G$ is measured directly by pulling the sample out of the coil at the lowest
temperature \cite{Prozorov2006}. The susceptibility of a rectangular superconducting slab in
the Meissner state can be written in terms of $\lambda$(T) and a characteristic dimension $R$, as $4\pi\chi\left( T\right) =\lambda/R\tanh{\left( R/\lambda\right) }-1$ \cite{Prozorov2006}.

Microscopic, thermodynamic and transport measurements of the Ba(Fe$_{1-x}$Co$_{x}$)$_{2}$ crystals used in this study have shown that in this particular system, superconductivity coexists with the orthorhombic phase in the
underdoped regime \cite{Nini2008}. Our TDR measurements reveal similar features. Fig.~\ref{fig1}(a) shows TDR frequency shifts as a function of temperature for scans running from below T$_{c}$ to $\approx$ 120 K for two
samples with $x=$~0.038 and 0.058. The data for the $x=$~0.058 sample has been divided by a factor of 5 for clarity. In the normal state, the magnetic penetration depth is limited by the skin depth, which depends on the
normal-state resistivity. The overall variation of $\Delta f$ over the transition region is about 20 Hz , which corresponds to a variation of about 45 nm in the skin depth. This should be compared to the 13300 Hz change
corresponding to the superconducting transition of the sample. Detection of the high temperature transition, which was determined by eye as the minimum in $\Delta f (T)$ (see Fig.~\ref{fig1}), as well as the superconducting $T_{c}$, which was determined by eye as the onset, are in excellent agreement with transport measurements \cite{Nini2008}, as shown in Fig.~\ref{fig1}(b). Fig.~\ref{fig2} shows the {\it rf} susceptibility constructed from the TDR frequency shifts in Ba(Fe$_{1-x}$Co$_{x}$)$_{2}$As$_{2}$ for $x=$~0.038, 0.047, 0.058, 0.074 and 0.10, which cover the range from underdoped to overdoped. Optimal doping for this series occurs for a concentration between $x=$~0.058 and 0.074.

\begin{figure}[tb]
\includegraphics[width=.95\columnwidth]{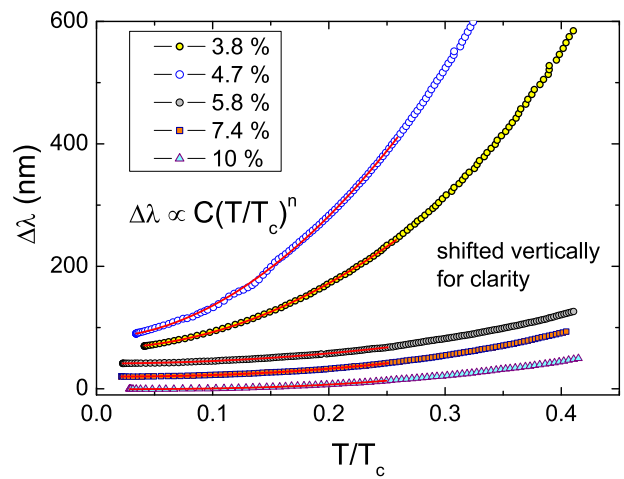}
\caption{(Color online) Low-temperature behavior of $\Delta \lambda(T)$ for all
studied concentrations. Solid lines are the fits to $\Delta \lambda \propto C(T/T_{c})^{n}$
with $C$ and $n$ as free parameters.}
\label{fig3}
\end{figure}

The low-temperature variation of the penetration depth is examined in Fig.~\ref{fig3}.
For all superconducting samples we observe a power law $\Delta\lambda(T) \propto C
T^{n}$. The fitted exponent $n$ varies from $n = 2 \pm 0.1$ for underdoped
samples to $n=2.5\pm0.1$ for the overdoped samples. If the superconducting
density itself follows a power law with a given $n$, then $C=f_s(c/\omega_{p}) S$,
where $f_{s}$ is the superconducting fraction at zero temperature, $S$ is
defined by the fraction of the Fermi surface that is gapless (which may reflect a
multigap character of the superconductivity, possible nodal structure, unitary impurity
scattering strength, etc) and $\omega_{p}$ is the plasma frequency.

\begin{figure}[tb]
\includegraphics[width=.95\columnwidth]{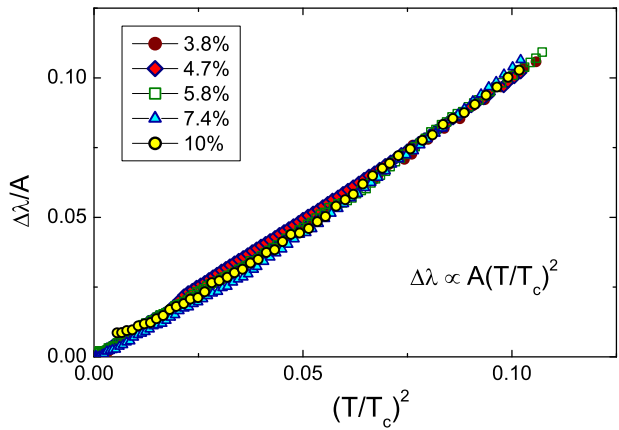}
\caption{(Color online) Scaled $\Delta \lambda(T)/A$ vs $(T/T_c)^2$ where $A$ is
obtained from the fitting, $\Delta \lambda \propto A(T/T_{c})^2$.}
\label{fig4}
\end{figure}

To examine how close the overall power-law behavior is to the quadratic one, we plot in
Fig.\ref{fig4} $\Delta\lambda((T/T_c)^2)$ scaled by the prefactor $A$ obtained
from the fitting of the data to $\Delta\lambda(T) \propto A(T/T_{c})^2$ at low
temperatures (below 0.3 $T_c$) with only one free parameter, $A$. At a gross level, all samples follow the $\lambda(T) \sim T^{2}$ behavior rather well. 

\begin{figure}[tb]
\includegraphics[width=.95\columnwidth]{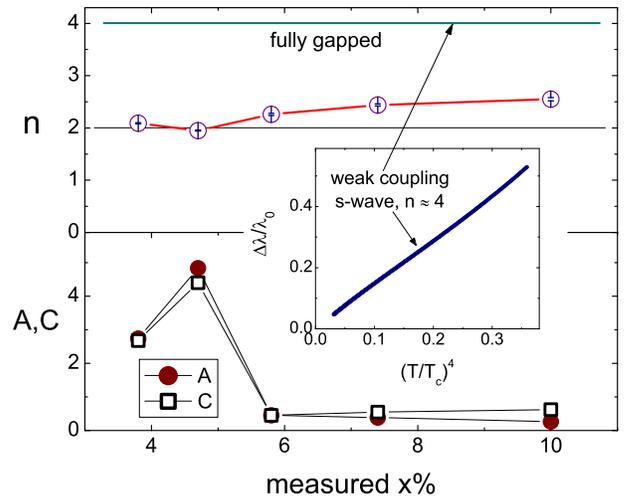}
\caption{(Color online) Doping dependence of the exponent $n$ (upper panel)
and of the fitting prefactors $A$ and $C$ (lower panel). The inset shows that the exponential
behavior described by Eq.~\protect\ref{eq1} is well represented by the power
law with $n \approx4$.}
\label{fig5}
\end{figure}

To summarize the observed power-law behavior, the upper panel of Fig.~\ref{fig5} shows the exponent $n$ as obtained from the best fit with two free parameters, $\Delta\lambda(T) \propto CT^{n}$. The lower panel of Fig.~\ref{fig5} shows the prefactor $C$ obtained from the above fit as well as a prefactor obtained by fitting to a pure quadratic behavior, $\Delta\lambda(T) \propto AT^{2}$. There is a clear change of regime at $x \sim 0.06$: (i) at lower $x$ the coexistence of antiferromagnetism and orthorhombicity with superconductivity has been inferred \cite{Nini2008} (also, see Fig.~\ref{fig1}), (ii) at $x \sim 0.06$, $n$ changes from $2.0 \pm 0.1$ to $2.4 \pm 0.1$ and (iii) the coefficient $C$ suddenly drops by an order of magnitude (Fig.~\ref{fig5}).

Let us now discuss the implications of these observations. If the Fermi
surface (FS) is fully gapped, the penetration depth of a homogeneous and
clean superconductor in the local limit exhibits the well-known exponential temperature
dependence:
\begin{equation}
\Delta \lambda \left( T\right) =\lambda \left( 0\right) \sqrt{\frac{\pi
\Delta _{\min }}{2k_{B}T}}\exp {\left( -\frac{\Delta _{\min }}{k_{B}T}%
\right) },  \label{eq1}
\end{equation}%
\noindent where $\Delta _{\min }$ is the minimal value of the gap. This holds roughly
until $k_{B}T\lesssim \Delta _{\min }/5$. Our data are not consistent with
this behavior. In what follows we comment on the viability of various scenarios that
yield power-law behavior of $\lambda(T)$. If a material is clean, in the local
limit, anisotropic pairing with line or point nodes of the pairing gap leads to a linear or quadratic temperature dependence of \ $\Delta \lambda \left( T\right) $, respectively. Thus, the most direct interpretation of our
data would be in terms of point nodes of the gap \cite{Einzel1986,Izawa2003}, 
as for example in PrOs$_4$Sb$_{12}$ \cite{Izawa2003}.
However, this is only correct if the system is clean and in the local limit. 
$\Delta \lambda \left( T\right)\propto T^{2}$ is consistent with line nodes \cite{note1} of the gap if one includes scattering by impurities \cite{Hirschfeld1993} or nonlocal corrections \cite{Leggett}. Unitary impurity scattering creates a state with a quadratic behavior below some characteristic temperature \cite{Hirschfeld1993} $k_{B}T^{\ast }\sim \Gamma $, where $\Gamma $ is the impurity scattering rate. In our case the requirement would be $\Gamma \gtrsim k_{B}T_{c}/3.$ This explanation requires relatively strong impurity
scattering \cite{note2} with a substantial unitary component, consistent with the
fact that our samples are doped in the active plane, but implying that in
clean samples, i.e. samples doped away from the Fe-As planes, a linear
behavior should be restored. A quadratic $T$ dependence of $\Delta \lambda $
may also be the result of strong impurity scattering in a pairing state that
is fully gapped in the clean limit. The exponential behavior of Eq.\ref{eq1}
transforms to quadratic if the gap is driven into a gapless or near-gapless
regime by impurity scattering. As in Ref. \cite{Parker2008}, it would
require a scattering rate $\Gamma $ of the order of the smallest gap $\Delta
_{\min }$ and a relatively fine balance between the unitary and Born
scattering regimes. In clean samples, i.e. those doped away from the Fe-As
planes, an exponential, or respectively, linear behavior should be restored.
Indeed, the existing data for the 1111 system \cite{Martin2008,Malone2008,Tonyunp} surprisingly do not show a quadratic dependence; namely, an exponential one in the As-based compounds \cite{Martin2008,Malone2008} but a linear one in the P-based system \cite{Tonyunp}.  Furthermore, a problem with this explanation is that we do
not see any systematic dependence of the $\Delta \lambda (T)$ characteristics on impurity concentration, but rather an abrupt change of regime as we cross the structural/magnetic phase transition. 

Another mechanism that may transform the linear behavior in a state with line nodes into a quadratic one  at $T<T^{\ast }\simeq \Delta(0)\xi _{0}/\lambda (0)$ is due to nonlocal effects \cite{Leggett},  where $\xi_{0}$ is the coherence length. In the Fe-pnictides, however, $T^{\ast }$ would be less than 1 K.

The power-law behavior of $\Delta \lambda $ may also be a consequence of the material being inhomogeneous. While the observation of a smearing of the jump in heat capacity at $T_{c}$ in under- and over- doped samples \cite{Nini2008} may be considered a hint for such a scenario, we do see very homogeneous Meissner screening in magneto-optical measurements. Also the jump in $C_p$ is doping dependent and it is unlikely that any inhomogeneity can explain the universal behavior shown in Fig.~\ref{fig3} for all concentrations. 

In view of this discussion, it is tempting to look for an explanation on a
phenomenological level that would not rely on impurity scattering as a
crucial element changing the functional dependence of $\lambda $. Indeed, any
excitation coupled with electrons with an energy larger than $\sim 2\pi T$
is pair-breaking, including regular phonons. Moreover, for an $s_{\pm }$ or
a $d-$wave state \textit{even} phonons with arbitrarily small energies can be pairbreaking. The same holds for the coupling to other collective bosonic
modes, such as antiferromagnetic spin fluctuations.  Since thermally
excited bosons are needed for pairbreaking, the scattering rate $\Gamma $
becomes $T$-dependent \cite{Abanov01}. In the case of line nodes, where a $T$
-independent $\Gamma $ yields an exponent $n=2$,  strong scattering off of the
thermally excited bosons would always yield a smaller exponent.  Given the
special role that the AFM critical point seems to play, the possibility exists
that the pair breaking fluctuations are associated with an intermediate range
dynamic ordering, like the dynamic domains speculated in Ref. \cite{MazinJohannes}. These will have very small energy and a potential to be strong scatterers. A clear derivation of the exponent $n$ that results from such a picture is still missing.

To summarize, we have measured the temperature dependence of the penetration
depth in single crystal Co-doped BaFe$_{2}$As$_{2}$. The main observations
are: (1) the superconducting density in Ba(Fe$_{1-x}$Co$_{x}$)$_{2}$ changes quadratically with temperature to at least $T_{c}/3$ and so exponential or linear behavior can be safely excluded; (2) there is a sharp change in the $T $ dependence of the penetration depth (and probably in the absolute value at zero temperature), which occurs near the same concentration at which the magnetic ordering in the normal state disappears; (3) there is no visible sample quality effect on the exponent and amplitude of the temperature dependent part of the penetration depth.

The observed behavior is compatible with neither fully gapped nor
line node superconductivity; ``accidental`` point nodes within the $s_{\pm}$ model can be excluded based on weak concentration dependence of the power-law exponent, but could still be considered for alternative pairing mechanisms. Impurity-driven quadratic behavior is possible, but seems somewhat problematic even for a line-nodal state, and especially for a nodeless state, given no visible dependence on the impurity concentration. The possible influence of the proximity to antiferromagnetic ordering suggest an intriguing interpretation in terms of
pair-breaking defects of magnetic nature whose concentration is controlled by the temperature and not by Co concentration.

We thank A.~A.~Golubov, O.~V.~Dolgov, D.~Parker, A.~V.~Chubukov,
B.~A.~Bernevig and A.~Carrington for useful discussions. Work at the Ames
Laboratory was supported by the Department of Energy-Basic Energy Sciences
under Contract No. DE-AC02-07CH11358. R. P. acknowledges support from Alfred
P. Sloan Foundation.


\begin{thebibliography}{99}
\bibitem{note2} It might be reasonable to assume that a $T^{2}$ behavior in
a nodal superconducting state will be less sensitive to the specifics of the
impurity scattering than than an $s_{\pm}$ state, but this is a quantitative
issue that needs to be resolved by actual calculations.
\end{thebibliography}
\end{document}